\documentstyle[prd,aps,preprint,tighten,epsfig]{revtex}

\begin{document}

\draft

\title{A Novel Parametrization of $\tau$-lepton Dominance and
Simplified One-loop Renormalization-group Equations of Neutrino
Mixing Angles and CP-violating Phases}
\author{\bf Zhi-zhong Xing}
\address{CCAST (World Laboratory), P.O. Box 8730, Beijing 100080,
China \\
and Institute of High Energy Physics, Chinese Academy of Sciences,
\\
P.O. Box 918, Beijing 100049, China
\footnote{Mailing address} \\
({\it Electronic address: xingzz@mail.ihep.ac.cn}) } \maketitle

\begin{abstract}
We point out that the $\tau$-lepton dominance in the one-loop
renormalization-group (RG) equations of neutrino mixing quantities
allows us to set a criterion to choose the most suitable
parametrization of the lepton flavor mixing matrix $U$: its
elements $U^{}_{3i}$ (for $i=1,2,3$) should be as simple as
possible. Such a novel parametrization is quite different from the
``standard" one used in the literature and can lead to greatly
simplified RG equations for three mixing angles and the physical
CP-violating phase(s) in the standard model or its minimal
supersymmetric extension, no matter whether neutrinos are Dirac or
Majorana particles. Some important features of our analytical
results and their phenomenological consequences are also
discussed.
\end{abstract}


\newpage

\section{Introduction}

The fact that neutrinos have non-vanishing masses is a clean
signal of new physics beyond the standard model (SM). To
understand the small neutrino mass-squared differences and the
large lepton flavor mixing angles observed in solar and
atmospheric neutrino oscillation experiments \cite{SNO,SK,KM,K2K},
many models based on either new flavor symmetries or some
unspecified interactions have been proposed at some superhigh
energy scales \cite{Review}. Their phenomenological consequences
at low energy scales can be confronted with current experimental
data, after the renormalization-group (RG) effects on those
neutrino mixing parameters are properly taken into account. Such
radiative corrections can be very significant in some cases, for
instance, when the masses of three light neutrinos are nearly
degenerate or the value of $\tan\beta$ is very large in the
minimal supersymmetric standard model (MSSM).

An elegant idea to explain the smallness of left-handed neutrino
masses is to introduce very heavy right-handed neutrinos and
lepton number violation into the SM or MSSM and to make use of the
famous seesaw mechanism \cite{SS}. Below the seesaw scale, where
heavy Majorana neutrinos become decoupled, the effective neutrino
coupling matrix $\kappa$ obeys the following one-loop RG equation
\cite{RGE}:
\begin{equation}
16\pi^2 \frac{{\rm d}\kappa}{{\rm d}t} = \alpha^{}_{\rm M} \kappa
+ C \left [ \left (Y^{}_lY^\dagger_l \right ) \kappa + \kappa
\left (Y^{}_l Y^\dagger_l \right )^T \right ] \; ,
\end{equation}
where $t\equiv \ln (\mu/\Lambda_{\rm SS})$ with $\mu$ being an
arbitrary renormalization scale between the electroweak scale
$\Lambda_{\rm EW} \sim 10^2$ GeV and the typical seesaw scale
$\Lambda_{\rm SS} \sim 10^{10\cdot\cdot\cdot 14}$ GeV, and
$Y^{}_l$ is the charged-lepton Yukawa coupling matrix. In the SM,
$C=-1.5$ and $\alpha^{}_{\rm M} \approx -3g^2_2 + 6y^2_t +
\lambda$; and in the MSSM, $C=1$ and $\alpha^{}_{\rm M} \approx
-1.2g^2_1 - 6g^2_2 + 6 y^2_t$, where $g^{}_1$ and $g^{}_2$ denote
the gauge couplings, $y^{}_t$ stands for the top-quark Yukawa
coupling, and $\lambda$ is the Higgs self-coupling in the SM.

There are also some good reasons to speculate that massive
neutrinos might be the Dirac particles \cite{Dirac}. In this case,
the Dirac neutrino Yukawa coupling matrix $Y^{}_\nu$ must be
extremely suppressed in magnitude, so as to reproduce the light
neutrino masses of ${\cal O}(1)$ eV or smaller at the electroweak
scale. $Y^{}_\nu$ can run from a superhigh energy scale down to
$\Lambda_{\rm EW}$ via the one-loop RG equation
\begin{equation}
16\pi^2 \frac{{\rm d} \omega}{{\rm d}t} = 2 \alpha^{}_{\rm D}
\omega + C \left [ \left (Y^{}_lY^\dagger_l \right ) \omega +
\omega \left (Y^{}_l Y^\dagger_l \right )\right ] \; ,
\end{equation}
where $\omega \equiv Y^{}_\nu Y^\dagger_\nu$, $\alpha^{}_{\rm D}
\approx -0.45g^2_1 - 2.25g^2_2 + 3y^2_t$ in the SM or
$\alpha^{}_{\rm D} \approx -0.6g^2_1 - 3g^2_2 + 3y^2_t$ in the
MSSM \cite{Dirac}. In obtaining Eq. (2), we have safely neglected
those tiny terms of ${\cal O}(\omega^2)$.

Eq. (1) or (2) allows us to derive the explicit RG equations for
all neutrino mass and mixing parameters in the flavor basis where
$Y^{}_l$ is diagonal and real (positive). In this basis, we have
$\kappa = {\cal V}^{}_{\rm M} \overline{\kappa} {\cal V}^T_{\rm
M}$ with $\overline{\kappa} = {\rm Diag}\{\kappa^{}_1,
\kappa^{}_2, \kappa^{}_3\}$ for Majorana neutrinos; or $\omega =
{\cal V}^{}_{\rm D} \overline{\omega} {\cal V}^\dagger_{\rm D}$
with $\overline{\omega} = {\rm Diag} \{ y^2_1, y^2_2, y^2_3 \}$
for Dirac neutrinos. ${\cal V}^{}_{\rm M}$ or ${\cal V}_{\rm D}$
is just the lepton flavor mixing matrix. At $\Lambda_{\rm EW}$,
Majorana neutrino masses are given by $m^{}_i = v^2 \kappa^{}_i$
(SM) or $m^{}_i = v^2 \kappa^{}_i \sin^2\beta$ (MSSM), while
Dirac neutrino masses are given by $m^{}_i = v y^{}_i$ (SM) or
$m^{}_i = v y^{}_i \sin\beta$ (MSSM) with $v \approx 174$ GeV.

Note that ${\cal V}^{}_{\rm M}$ (or ${\cal V}^{}_{\rm D}$) can be
parametrized in terms of three mixing angles and a few
CP-violating phases. Their RG equations consist of the
flavor-dependent contributions from $Y^{}_l Y^\dagger_l$. Because
of $y^2_e \ll y^2_\mu \ll y^2_\tau$, where $y^{}_e$, $y^{}_\mu$
and $y^{}_\tau$ correspond to the electron, muon and tau Yukawa
couplings, we only need to take account of the dominant
$\tau$-lepton contribution to those one-loop RG equations of
neutrino mixing angles and CP-violating phases in an excellent
approximation. A careful analysis shows that the $\tau$-dominance
is closely associated with the matrix elements $\left ({\cal
V}^{}_{\rm M} \right )_{3i}$ or $\left ({\cal V}^{}_{\rm D} \right
)_{3i}$ (for $i=1,2,3$). This important observation implies that
very concise RG equations can be obtained for those flavor mixing
and CP-violating parameters, if ${\cal V}^{}_{\rm M}$ (or ${\cal
V}^{}_{\rm D}$) is parametrized in such a way that its elements
$\left ({\cal V}^{}_{\rm M} \right )_{3i}$ (or $\left ({\cal
V}^{}_{\rm D} \right )_{3i}$) are as simple as possible. One may
then make use of this criterion to choose the most suitable
parametrization of ${\cal V}^{}_{\rm M}$ or ${\cal V}^{}_{\rm D}$
in deriving the one-loop RG equations.

We find that the so-called ``standard" parametrization (advocated
by the Particle Data Group \cite{PDG}), which has extensively been
used in describing lepton flavor mixing, does not satisfy the
above criterion. Instead, the parametrization recommended in Ref.
\cite{FX97} fulfills our present requirement
\footnote{This parametrization may naturally arise from the
parallel (and probably hierarchical) textures of charged-lepton
and neutrino mass matrices \cite{FX97}. It is phenomenologically
possible to obtain $\theta^{}_l \approx \arctan \left
(\sqrt{m^{}_e/m^{}_\mu} ~\right ) \approx 4^\circ$ \cite{FX96} and
a suggestive relationship between $\theta^{}_\nu$ and
$m^{}_1/m^{}_2$ \cite{FX05}.}:
\begin{eqnarray}
U & = & \left ( \matrix{ c^{}_l & s^{}_l   & 0 \cr -s^{}_l    &
c^{}_l   & 0 \cr 0   & 0 & 1 \cr } \right )  \left ( \matrix{
e^{-i\phi}  & 0 & 0 \cr 0   & c & s \cr 0   & -s    & c \cr }
\right )  \left ( \matrix{ c^{}_{\nu} & -s^{}_{\nu}  & 0 \cr
s^{}_{\nu} & c^{}_{\nu}   & 0 \cr
0   & 0 & 1 \cr } \right )  \nonumber \\ \nonumber \\
& = & \left ( \matrix{ s^{}_l s^{}_{\nu} c + c^{}_l c^{}_{\nu}
e^{-i\phi} & s^{}_l c^{}_{\nu} c - c^{}_l s^{}_{\nu} e^{-i\phi} &
s^{}_l s \cr c^{}_l s^{}_{\nu} c - s^{}_l c^{}_{\nu} e^{-i\phi} &
c^{}_l c^{}_{\nu} c + s^{}_l s^{}_{\nu} e^{-i\phi} & c^{}_l s \cr
- s^{}_{\nu} s   & - c^{}_{\nu} s   & c \cr } \right ) \; ,
\end{eqnarray}
where $c^{}_l \equiv \cos\theta^{}_l$, $s^{~}_l \equiv
\sin\theta_l$, $c^{}_\nu \equiv \cos\theta^{}_\nu$, $s^{}_{\nu}
\equiv \sin\theta^{}_{\nu}$, $c \equiv \cos\theta$ and $s \equiv
\sin\theta$. In general, we have ${\cal V}^{}_{\rm M} = Q^{}_{\rm
M} U P^{}_{\rm M}$ for Majorana neutrinos or ${\cal V}^{}_{\rm D}
= Q^{}_{\rm D}UP^{}_{\rm D}$ for Dirac neutrinos, where $P^{}_{\rm
M}$ (or $P^{}_{\rm D}$) and $Q^{}_{\rm M}$ (or $Q^{}_{\rm D}$) are
two diagonal phase matrices. It is clear that $U^{}_{3i}$ (for
$i=1,2,3$) shown in Eq. (3) are simple enough to describe the
$\tau$-dominant terms in those one-loop RG equations of
$\theta^{}_l$, $\theta^{}_\nu$, $\theta$ and $\phi$ (as well as
two Majorana phases of ${\cal V}^{}_{\rm M}$ coming from
$P^{}_{\rm M}$). In the approximation that solar and atmospheric
neutrino oscillations are nearly decoupled \cite{Vissani}, three
mixing angles of $U$ can simply be related to those of solar,
atmospheric and CHOOZ neutrino oscillations \cite{SNO,SK,KM}:
$\theta^{}_{\rm sun} \approx \theta^{}_\nu$, $\theta^{}_{\rm atm}
\approx \theta$ and $\theta^{}_{\rm chz} \approx \theta^{}_l
\sin\theta$. Hence our parametrization is also a convenient option
to describe current neutrino oscillation data.

The main purpose of this paper is to show that Eq. (3) is actually
a novel parametrization of $\tau$-dominance in the one-loop RG
equations of neutrino mixing angles and CP-violating phases.
Compared with the ``standard" parametrization used in the
literature, Eq. (3) leads to greatly simplified results for
relevant RG equations. The latter can therefore allow us to
understand the RG running behaviors of lepton flavor mixing
parameters in a much simpler and more transparent way, which is of
course useful for model building at a superhigh energy scale to
explore possible flavor symmetries or flavor dynamics responsible
for the origin of neutrino masses and CP violation.

In section II, we use Eqs. (1) and (3) to derive the one-loop RG
equations of three mixing angles and three CP-violating phases for
Majorana neutrinos. Section III is devoted to the one-loop RG
equations of three mixing angles and one CP-violating phase for
Dirac neutrinos, and to a brief comparison between the Jarlskog
invariants of CP violation in Dirac and Majorana cases. Some
concluding remarks are given in section IV.

\section{RG equations for Majorana neutrinos}

The general strategy and tactics about how to derive the one-loop
RG equations for Majorana neutrino mixing parameters have been
outlined in Refs. \cite{Casas,Antusch,Antusch2,Mei}. To be
specific, we take $P^{}_{\rm M} = {\rm Diag} \left \{ e^{i\rho},
e^{i\sigma}, 1 \right \}$ and $Q^{}_{\rm M} = {\rm Diag} \left \{
e^{i\phi^{}_1}, e^{i\phi^{}_2}, e^{i\phi^{}_3} \right \}$. The
phase parameters $\rho$ and $\sigma$ are physical and referred to
as the Majorana phases. The phase parameters $\phi^{}_i$ (for
$i=1,2,3$) are unphysical, but they have their own RG evolution.
Following the procedure described in Ref. \cite{Casas} and taking
the $\tau$-dominance approximation, we obtain the RG equations of
$\kappa^{}_i$ (for $i=1,2,3$) from Eq. (1):
\begin{equation}
\dot{\kappa}^{}_i \; = \; \frac{\kappa^{}_i}{16\pi^2} \left (
\alpha^{}_{\rm M} + 2 C y^2_\tau \left | U^{}_{3i} \right |^2
\right ) \; ,
\end{equation}
where $\dot{\kappa}^{}_i \equiv {\rm d}\kappa^{}_i/{\rm d}t$. In
addition, the quantities $\rho$, $\sigma$, $\phi^{}_i$ and
$U^{}_{ij}$ (for $i,j =1,2,3$) satisfy the following equations:
\begin{eqnarray}
\sum^3_{j=1} \left [ U^*_{j1} \left (i \dot{U}^{}_{j1} - U^{}_{j1}
\dot{\phi}^{}_j \right ) \right ] & = & \dot{\rho} \; ,
\nonumber \\
\sum^3_{j=1} \left [ U^*_{j2} \left (i \dot{U}^{}_{j2} - U^{}_{j2}
\dot{\phi}^{}_j \right ) \right ] & = & \dot{\sigma} \; ,
\nonumber \\
\sum^3_{j=1} \left [ U^*_{j3} \left (i \dot{U}^{}_{j3} - U^{}_{j3}
\dot{\phi}^{}_j \right ) \right ] & = & 0 \; ;
\end{eqnarray}
and
\begin{eqnarray}
\sum^3_{j=1} \left [ U^*_{j1} \left (\dot{U}^{}_{j2} + i U^{}_{j2}
\dot{\phi}^{}_j \right ) \right ] & = & -
\frac{Cy^2_\tau}{16\pi^2} ~ e^{i\left (\rho -\sigma \right )}
\left [ \zeta^{-1}_{12} {\rm Re} \left ( U^*_{31} U^{}_{32}
e^{i\left (\sigma -\rho \right )} \right ) + i \zeta^{}_{12} {\rm
Im} \left ( U^*_{31} U^{}_{32} e^{i\left (\sigma -\rho \right )}
\right ) \right ] \; ,
\nonumber \\
\sum^3_{j=1} \left [ U^*_{j1} \left (\dot{U}^{}_{j3} + i U^{}_{j3}
\dot{\phi}^{}_j \right ) \right ] & = & -
\frac{Cy^2_\tau}{16\pi^2} ~ e^{i \rho} \left [ \zeta^{-1}_{13}
{\rm Re} \left ( U^*_{31} U^{}_{33} e^{-i\rho} \right ) + i
\zeta^{}_{13} {\rm Im} \left ( U^*_{31} U^{}_{33} e^{-i\rho}
\right ) \right ] \; ,
\nonumber \\
\sum^3_{j=1} \left [ U^*_{j2} \left (\dot{U}^{}_{j3} + i U^{}_{j3}
\dot{\phi}^{}_j \right ) \right ] & = & -
\frac{Cy^2_\tau}{16\pi^2} ~ e^{i \sigma} \left [ \zeta^{-1}_{23}
{\rm Re} \left ( U^*_{32} U^{}_{33} e^{-i\sigma} \right ) + i
\zeta^{}_{23} {\rm Im} \left ( U^*_{32} U^{}_{33} e^{-i\sigma}
\right ) \right ] \; ,
\end{eqnarray}
where $\zeta^{}_{ij} \equiv \left (\kappa^{}_i - \kappa^{}_j
\right )/\left (\kappa^{}_i + \kappa^{}_j \right )$. Obviously,
those $y^2_\tau$-associated terms only consist of the matrix
elements $U^{}_{3i}$ (for $i=1,2,3$). If a parametrization of $U$
assures $U^{}_{3i}$ to be as simple as possible, then the
resultant RG equations of relevant neutrino mixing angles and
CP-violating phases will be as concise as possible. One can see
that the parametrization of $U$ given in Eq. (3) just accords with
such a criterion, while the ``standard" parametrization advocated
in Ref. \cite{PDG} and used in many papers (e.g., Refs.
\cite{Casas,Antusch,Antusch2,Mei,Luo1}) does not satisfy this
requirement.

Combining Eq. (3) with Eqs. (4), (5) and (6), we arrive at
\begin{eqnarray}
\dot{\kappa}^{}_1 & = & \frac{\kappa^{}_1}{16\pi^2} \left (
\alpha^{}_{\rm M} + 2 C y^2_\tau s^2_\nu s^2 \right ) \; ,
\nonumber \\
\dot{\kappa}^{}_2 & = & \frac{\kappa^{}_2}{16\pi^2} \left (
\alpha^{}_{\rm M} + 2 C y^2_\tau c^2_\nu s^2 \right ) \; ,
\nonumber \\
\dot{\kappa}^{}_3 & = & \frac{\kappa^{}_3}{16\pi^2} \left (
\alpha^{}_{\rm M} + 2 C y^2_\tau c^2 \right ) \; ;
\end{eqnarray}
and
\begin{eqnarray}
\dot{\theta}^{}_l & = & \frac{C y^2_\tau}{16\pi^2} ~ c^{}_\nu
s^{}_\nu c \left [ \left ( \zeta^{-1}_{13} c^{}_\rho c^{}_{(\rho
-\phi)} + \zeta^{}_{13} s^{}_\rho s^{}_{(\rho - \phi)} \right ) -
\left ( \zeta^{-1}_{23} c^{}_\sigma c^{}_{(\sigma -\phi)} +
\zeta^{}_{23} s^{}_\sigma s^{}_{(\sigma - \phi)} \right ) \right ]
\; ,
\nonumber \\
\dot{\theta}^{}_\nu & = & \frac{C y^2_\tau}{16\pi^2} ~ c^{}_\nu
s^{}_\nu \left [ s^2 \left ( \zeta^{-1}_{12} c^2_{(\sigma -\rho)}
+ \zeta^{}_{12} s^2_{(\sigma -\rho)} \right ) + c^2 \left (
\zeta^{-1}_{13} c^2_\rho + \zeta^{}_{13} s^2_\rho \right ) - c^2
\left ( \zeta^{-1}_{23} c^2_\sigma + \zeta^{}_{23} s^2_\sigma
\right ) \right ] \; ,
\nonumber \\
\dot{\theta} \; & = & \frac{C y^2_\tau}{16\pi^2} ~ c s \left [
s^2_\nu \left ( \zeta^{-1}_{13} c^2_\rho + \zeta^{}_{13} s^2_\rho
\right ) + c^2_\nu \left ( \zeta^{-1}_{23} c^2_\sigma +
\zeta^{}_{23} s^2_\sigma \right ) \right ] \; ;
\end{eqnarray}
as well as
\begin{eqnarray}
\dot{\phi} & = & \frac{C y^2_\tau}{16\pi^2} \left [ \left ( c^2_l
-s^2_l \right ) c^{-1}_l s^{-1}_l c^{}_\nu s^{}_\nu c \left (
\zeta^{-1}_{13} c^{}_\rho s^{}_{(\rho -\phi)} - \zeta^{}_{13}
s^{}_\rho c^{}_{(\rho -\phi)} - \zeta^{-1}_{23} c^{}_\sigma
s^{}_{(\sigma -\phi)} + \zeta^{}_{23} s^{}_\sigma c^{}_{(\sigma
-\phi)} \right ) \right .
\nonumber \\
& & \left . ~~~~~ + \widehat{\zeta}^{}_{12} s^2 c^{}_{(\sigma
-\rho)} s^{}_{(\sigma -\rho)} + \widehat{\zeta}^{}_{13} \left
(s^2_\nu - c^2_\nu c^2 \right ) c^{}_\rho s^{}_\rho +
\widehat{\zeta}^{}_{23} \left (c^2_\nu - s^2_\nu c^2 \right )
c^{}_\sigma s^{}_\sigma \right ] \; ,
\nonumber \\
\dot{\rho} & = & \frac{C y^2_\tau}{16\pi^2} \left [
\widehat{\zeta}^{}_{12} c^2_\nu s^2 c^{}_{(\sigma -\rho)}
s^{}_{(\sigma -\rho)} + \widehat{\zeta}^{}_{13} \left (s^2_\nu s^2
- c^2 \right ) c^{}_\rho s^{}_\rho + \widehat{\zeta}^{}_{23}
c^2_\nu s^2 c^{}_\sigma s^{}_\sigma \right ] \; ,
\nonumber \\
\dot{\sigma} & = & \frac{C y^2_\tau}{16\pi^2} \left [
\widehat{\zeta}^{}_{12} s^2_\nu s^2 c^{}_{(\sigma -\rho)}
s^{}_{(\sigma -\rho)} + \widehat{\zeta}^{}_{13} s^2_\nu s^2
c^{}_\rho s^{}_\rho + \widehat{\zeta}^{}_{23} \left ( c^2_\nu s^2
- c^2 \right ) c^{}_\sigma s^{}_\sigma \right ] \; ,
\end{eqnarray}
where $\widehat{\zeta}^{}_{ij} \equiv \zeta^{-1}_{ij} -
\zeta^{}_{ij} = 4 \kappa^{}_i \kappa^{}_j/\left ( \kappa^2_i -
\kappa^2_j \right )$, $c^{}_a \equiv \cos a$ and $s^{}_a \equiv
\sin a$ (for $a = \rho$, $\sigma$, $\sigma -\rho$, $\rho -\phi$ or
$\sigma -\phi$). Comparing the RG equations of three mixing angles
and three CP-violating phases obtained in Eqs. (8) and (9) with
their counterparts given in Refs.
\cite{Casas,Antusch,Antusch2,Mei,Luo1}, which were derived by
using the ``standard" parametrization, we find that great
simplification and conciseness have been achieved for our present
analytical results.

As a by-product, the RG equations of three unphysical phases
$\phi^{}_i$ are listed below:
\begin{eqnarray}
\dot{\phi}^{}_1 & = & + \frac{Cy^2_\tau}{16\pi^2} \left [ c^{}_l
s^{-1}_l c^{}_\nu s^{}_\nu c \left ( \zeta^{-1}_{13} c^{}_\rho
s^{}_{(\rho - \phi)} - \zeta^{}_{13} s^{}_\rho c^{}_{(\rho -
\phi)} - \zeta^{-1}_{23} c^{}_\sigma s^{}_{(\sigma - \phi)} +
\zeta^{}_{23} s^{}_\sigma c^{}_{(\sigma - \phi)} \right ) \right .
\nonumber \\
& & \left . ~~~~~~~ + c^2 \left ( \widehat{\zeta}^{}_{13} s^2_\nu
c^{}_\rho s^{}_\rho + \widehat{\zeta}^{}_{23} c^2_\nu c^{}_\sigma
s^{}_\sigma \right ) \right ] \; ,
\nonumber \\
\dot{\phi}^{}_2 & = & - \frac{Cy^2_\tau}{16\pi^2} \left [ c^{-1}_l
s^{}_l c^{}_\nu s^{}_\nu c \left ( \zeta^{-1}_{13} c^{}_\rho
s^{}_{(\rho - \phi)} - \zeta^{}_{13} s^{}_\rho c^{}_{(\rho -
\phi)} - \zeta^{-1}_{23} c^{}_\sigma s^{}_{(\sigma - \phi)} +
\zeta^{}_{23} s^{}_\sigma c^{}_{(\sigma - \phi)} \right ) \right .
\nonumber \\
& & \left . ~~~~~~~ - c^2 \left ( \widehat{\zeta}^{}_{13} s^2_\nu
c^{}_\rho s^{}_\rho + \widehat{\zeta}^{}_{23} c^2_\nu c^{}_\sigma
s^{}_\sigma \right ) \right ] \; ,
\nonumber \\
\dot{\phi}^{}_3 & = & - \frac{Cy^2_\tau}{16\pi^2} \left [ s^2
\left ( \widehat{\zeta}^{}_{13} s^2_\nu c^{}_\rho s^{}_\rho +
\widehat{\zeta}^{}_{23} c^2_\nu c^{}_\sigma s^{}_\sigma \right )
\right ] \; .
\end{eqnarray}
It is easy to check that the relationship $\dot{\phi} = \dot{\rho}
+ \dot{\sigma} + \dot{\phi}^{}_1 + \dot{\phi}^{}_2 +
\dot{\phi}^{}_3$ holds. That is why $\phi^{}_i$ should not be
ignored in deriving the RG equations of other physical parameters,
although these three phases can finally be rotated away via
rephasing the charged-lepton fields.

Some qualitative comments on the basic features of Eqs. (7)--(10)
are in order.

(a) The RG running behaviors of three neutrino masses $m^{}_i$ (or
equivalently $\kappa^{}_i$) are essentially identical and
determined by $\alpha^{}_{\rm M}$ \cite{Antusch}, unless
$\tan\beta$ is large enough in the MSSM to make the
$y^2_\tau$-associated term is competitive with the $\alpha^{}_{\rm
M}$ term. In our phase convention, $\dot{\kappa}^{}_i$ or
$\dot{m}^{}_i$ (for $i=1,2,3$) are independent of the CP-violating
phase $\phi$.

(b) Among three mixing angles, only the derivative of
$\theta^{}_\nu$ contains a term proportional to $\zeta^{-1}_{12}$.
Note that $\zeta^{-1}_{ij} = - ( m^{}_i + m^{}_j )^2/\Delta
m^2_{ji}$ with $\Delta m^2_{ji} \equiv m^2_j - m^2_i$ holds, and
current solar and atmospheric neutrino oscillation data yield
$\Delta m^2_{21} \approx 8 \times 10^{-5} ~ {\rm eV}^2$ and $\left
| \Delta m^2_{32} \right | \approx \left | \Delta m^2_{31} \right
| \approx 2.5 \times 10^{-3} ~ {\rm eV}^2$ \cite{Vissani}. Thus
$\theta^{}_\nu$ is in general more sensitive to radiative
corrections than $\theta^{}_l$ and $\theta$. The RG running of
$\theta^{}_\nu$ can be suppressed through the fine-tuning of
$(\sigma -\rho)$. The smallest mixing angle $\theta^{}_l$ may get
radiative corrections even if its initial value is zero, thus it
can be radiatively generated from other mixing angles and
CP-violating phases.

(c) The RG running behavior of $\phi$ is quite different from
those of $\rho$ and $\sigma$, because it includes a peculiar term
proportional to $s^{-1}_l$. This term, which dominates
$\dot{\phi}$ when $\theta^{}_l$ is sufficiently small, becomes
divergent in the limit $\theta^{}_l \rightarrow 0$. Indeed, $\phi$
is not well-defined if $\theta^{}_l$ is exactly vanishing. But
both $\theta^{}_l$ and $\phi$ can be radiatively generated. We may
require that $\dot{\phi}$ should remain finite when $\theta^{}_l$
approaches zero, implying that the following necessary condition
can be extracted from the expression of $\dot{\phi}$ in Eq. (9):
\begin{equation}
\zeta^{-1}_{13} c^{}_\rho s^{}_{(\rho -\phi)} - \zeta^{}_{13}
s^{}_\rho c^{}_{(\rho -\phi)} - \zeta^{-1}_{23} c^{}_\sigma
s^{}_{(\sigma -\phi)} + \zeta^{}_{23} s^{}_\sigma c^{}_{(\sigma
-\phi)} \; = \; 0 \; .
\end{equation}
It turns out that
\begin{equation}
\tan\phi = \frac{\widehat{\zeta}^{}_{13} \sin 2\rho -
\widehat{\zeta}^{}_{23} \sin 2\sigma}{\left ( \zeta^{-1}_{13} +
\zeta^{}_{13} + \widehat{\zeta}^{}_{13} \cos 2\rho \right ) -
\left ( \zeta^{-1}_{23} + \zeta^{}_{23} + \widehat{\zeta}^{}_{23}
\cos 2\sigma \right )} \;
\end{equation}
holds, a result similar to the one obtained in Eq. (25) of Ref.
\cite{Antusch}. Note that the initial value of $\theta^{}_l$, if
it is exactly zero or extremely small, may immediately drive
$\phi$ to its {\it quasi-fixed point} (see Ref. \cite{Luo2} for a
relevant study of the quasi-fixed point in the ``standard"
parametrization of lepton flavor mixing). In this interesting
case, Eq. (12) can be used to understand the relationship between
$\phi$ and two Majorana phases $\rho$ and $\sigma$ at the
quasi-fixed point.

(d) On the other hand, the RG running behaviors of $\rho$ and
$\sigma$ are relatively mild in comparison with that of $\phi$. A
remarkable feature of $\dot{\rho}$ and $\dot{\sigma}$ is that they
will vanish, if both $\rho$ and $\sigma$ are initially vanishing.
This observation indicates that $\rho$ and $\sigma$ cannot
simultaneously be generated from $\phi$ via the one-loop RG
evolution. In contrast, a different conclusion was drawn in Ref.
\cite{Luo1}, where the ``standard" parametrization with a slightly
changed phase convention was utilized.

(e) As for three unphysical phases, $\phi^{}_2$ and $\phi^{}_3$
only have relatively mild RG running effects, while the running
behavior of $\phi^{}_1$ may be violent for sufficiently small
$\theta^{}_l$. A quasi-fixed point of $\phi^{}_1$ is also expected
in the limit $\theta^{}_l \rightarrow 0$ and under the
circumstance given by Eq. (11) or (12).

\section{RG equations for Dirac neutrinos}

Now let us derive the one-loop RG equations for Dirac neutrino
mixing parameters. To be specific, we take $P^{}_{\rm D} = {\rm
Diag} \left \{ e^{i\varphi^{}_1}, e^{i\varphi^{}_2},
e^{i\varphi^{}_3} \right \}$ and $Q^{}_{\rm D} = {\rm Diag} \left
\{ e^{i\alpha}, e^{i\beta}, 1 \right \}$. The phase matrix
$P^{}_{\rm D}$ can be cancelled in $\omega$, thus it does not take
part in the RG evolution. The phase parameters $\alpha$ and
$\beta$ are also unphysical, but they have their own RG running
behaviors. Following the procedure described in Refs.
\cite{Dirac,Zhang} and taking the $\tau$-dominance approximation,
we get the RG equations of $y^{}_i$ (for $i=1,2,3$) from Eq. (2):
\begin{equation}
\dot{y}^{}_i \; = \; \frac{y^{}_i}{16\pi^2} \left ( \alpha^{}_{\rm
D} + C y^2_\tau \left | U^{}_{3i} \right |^2 \right ) \; ,
\end{equation}
where $\dot{y}^{}_i \equiv {\rm d}y^{}_i/{\rm d}t$. On the other
hand, the quantities $\alpha$, $\beta$ and $U^{}_{ij}$ (for $i,j
=1,2,3$) satisfy the following equations:
\begin{eqnarray}
\sum^3_{j=1} \left (U^*_{j1} \dot{U}^{}_{j2} \right ) + i \left (
\dot{\alpha} U^*_{11} U^{}_{12} + \dot{\beta} U^*_{21} U^{}_{22}
\right ) & = & -\frac{Cy^2_\tau}{16\pi^2} ~ \xi^{}_{12} U^*_{31}
U^{}_{32} \; ,
\nonumber \\
\sum^3_{j=1} \left (U^*_{j1} \dot{U}^{}_{j3} \right ) + i \left (
\dot{\alpha} U^*_{11} U^{}_{13} + \dot{\beta} U^*_{21} U^{}_{23}
\right ) & = & -\frac{Cy^2_\tau}{16\pi^2} ~ \xi^{}_{13} U^*_{31}
U^{}_{33} \; ,
\nonumber \\
\sum^3_{j=1} \left (U^*_{j2} \dot{U}^{}_{j3} \right ) + i \left (
\dot{\alpha} U^*_{12} U^{}_{13} + \dot{\beta} U^*_{22} U^{}_{23}
\right ) & = & -\frac{Cy^2_\tau}{16\pi^2} ~ \xi^{}_{23} U^*_{32}
U^{}_{33} \; ,
\end{eqnarray}
where $\xi^{}_{ij} \equiv \left (y^2_i + y^2_j \right )/\left
(y^2_i - y^2_j \right )$. Again, the $y^2_\tau$-associated terms
in Eqs. (13) and (14) only contain $U^{}_{3i}$ (for $i=1,2,3$).
These RG equations can therefore be specified in a relatively
concise way, if the parametrization of $U$ shown in Eq. (3) is
taken into account.

Explicitly, the Yukawa coupling eigenvalues of three Dirac
neutrinos obey the one-loop RG equations
\begin{eqnarray}
\dot{y}^{}_1 & = & \frac{y^{}_1}{16\pi^2} \left ( \alpha^{}_{\rm
D} + C y^2_\tau s^2_\nu s^2 \right ) \; ,
\nonumber \\
\dot{y}^{}_2 & = & \frac{y^{}_2}{16\pi^2} \left ( \alpha^{}_{\rm
D} + C y^2_\tau c^2_\nu s^2 \right ) \; ,
\nonumber \\
\dot{y}^{}_3 & = & \frac{y^{}_3}{16\pi^2} \left ( \alpha^{}_{\rm
D} + 2 C y^2_\tau c^2 \right ) \; .
\end{eqnarray}
The RG equations of three neutrino mixing angles and one
(physical) CP-violating phase are given by
\begin{eqnarray}
\dot{\theta}^{}_l & = & + \frac{Cy^2_\tau}{16\pi^2} ~ c^{}_\nu
s^{}_\nu c c^{}_\phi \left ( \xi^{}_{13} - \xi^{}_{23} \right ) \;
,
\nonumber \\
\dot{\theta}^{}_\nu & = & + \frac{Cy^2_\tau}{16\pi^2} ~ c^{}_\nu
s^{}_\nu \left [ s^2 \xi^{}_{12} + c^2 \left ( \xi^{}_{13} -
\xi^{}_{23} \right ) \right ] \; ,
\nonumber \\
\dot{\theta} \; & = & + \frac{Cy^2_\tau}{16\pi^2} ~ c s \left (
s^2_\nu \xi^{}_{13} + c^2_\nu \xi^{}_{23} \right ) \; ,
\nonumber \\
\dot{\phi} \; & = & - \frac{Cy^2_\tau}{16\pi^2} \left ( c^2_l -
s^2_l \right ) c^{-1}_l s^{-1}_l c^{}_\nu s^{}_\nu c s^{}_\phi
\left ( \xi^{}_{13} - \xi^{}_{23} \right ) \; ,
\end{eqnarray}
where $c^{}_\phi \equiv \cos\phi$ and $s^{}_\phi \equiv \sin\phi$.
The RG equations of two unphysical phases $\alpha$ and $\beta$
read
\begin{eqnarray}
\dot{\alpha} & = & - \frac{Cy^2_\tau}{16\pi^2} ~ c^{}_l s^{-1}_l
c^{}_\nu s^{}_\nu c s^{}_\phi \left ( \xi^{}_{13} - \xi^{}_{23}
\right ) \; ,
\nonumber \\
\dot{\beta} & = & + \frac{Cy^2_\tau}{16\pi^2} ~ c^{-1}_l s^{}_l
c^{}_\nu s^{}_\nu c s^{}_\phi \left ( \xi^{}_{13} - \xi^{}_{23}
\right ) \; .
\end{eqnarray}
The relationship $\dot{\phi} = \dot{\alpha} + \dot{\beta}$ holds
obviously, implying that $\alpha$ and $\beta$ are not negligible
in deriving the RG equations of other physical parameters. One can
see that our analytical results are really concise, thanks to the
novel parametrization of $U$ that we have taken.

Some qualitative remarks on the main features of Eqs. (15), (16)
and (17) are in order.

(1) Like the Majorana case, the RG running behaviors of three
Dirac neutrino masses $m^{}_i$ (or equivalently $y^{}_i$) are
nearly identical and determined by $\alpha^{}_{\rm D}$
\cite{Dirac}, unless $\tan\beta$ is sufficiently large in the
MSSM. It is also worth mentioning that $\dot{y}^{}_i$ or
$\dot{m}^{}_i$ (for $i=1,2,3$) are independent of both the
CP-violating phase $\phi$ and the smallest mixing angle
$\theta^{}_l$ in our parametrization.

(2) The derivative of $\theta^{}_\nu$ consists of a term
proportional to $\xi^{}_{12} = - (m^2_1 + m^2_2)/\Delta m^2_{21}$.
Hence $\theta^{}_\nu$ is in general more sensitive to radiative
corrections than $\theta^{}_l$ and $\theta$, whose derivatives are
only dependent on $\xi_{13} = - (m^2_1 + m^2_3)/\Delta m^2_{31}$
and $\xi_{23} = - (m^2_2 + m^2_3)/\Delta m^2_{32}$. Given
$\theta^{}_\nu$ and $\theta$ at a specific energy scale, the
smallest mixing angle $\theta^{}_l$ can be radiatively generated
at another energy scale. In this case, however, it is impossible
to simultaneously generate the CP-violating phase $\phi$ (see Ref.
\cite{Dirac} for a similar conclusion in the ``standard"
parametrization of $U$). The reason is simply that $\phi$ can
always be rotated away when $\theta^{}_l$ is exactly vanishing,
and the proportionality relationship between $\dot{\phi}$ and
$\sin\phi$ forbids $\phi$ to be generated even when $\theta^{}_l$
becomes non-vanishing.

(3) Different from the Majorana case, there is no non-trivial {\it
quasi-fixed point} in the RG evolution of $\phi$ for Dirac
neutrinos. If $\dot{\phi}$ is required to keep finite when
$\theta^{}_l$ approaches zero, then $\phi$ itself must approach
zero or $\pi$, as indicated by Eq. (16). On the other hand,
$\dot{\theta}^{}_l \propto \cos\phi$ implies that the RG running
of $\theta^{}_l$ has a turning point characterized by $\phi =
\pi/2$ (i.e., $\dot{\theta}^{}_l$ flips its sign at this point).
Hence two interesting conclusions analogous to those drawn in Ref.
\cite{Dirac} can be achieved: first, $\theta^{}_l$ can never cross
zero if $\theta^{}_l \neq 0$ and $\sin\phi \neq 0$ hold at a
certain energy scale; second, CP will always be a good symmetry if
$\theta^{}_l = 0$ or $\sin\phi =0$ holds at a certain energy
scale.

(4) The RG running behavior of $\alpha$ is quite similar to that
of $\phi$, because $\dot{\phi} = \dot{\alpha} \left ( 1 -
\tan^2\theta^{}_l \right )$ holds. In addition, $\dot{\beta} = -
\dot{\alpha} \tan^2\theta^{}_l$ holds, implying that $\beta$ only
gets some relatively mild RG corrections.

Let us remark that the Jarlskog invariant of CP violation \cite{J}
takes the same form for Dirac and Majorana neutrinos: ${\cal J} =
c^{}_l s^{}_l c^{}_\nu s^{}_\nu c s^2 s^{}_\phi$. If neutrinos are
Dirac particles, the one-loop RG equation of ${\cal J}^{}_{\rm D}$
can be expressed as
\begin{equation}
\dot{\cal J}^{}_{\rm D} \; =\; \frac{Cy^2_\tau}{16\pi^2} ~ {\cal
J}^{}_{\rm D} \left [ \left (c^2_\nu - s^2_\nu \right ) s^2
\xi^{}_{12} + \left ( c^2 - s^2_\nu s^2 \right ) \xi^{}_{13} +
\left ( c^2 - c^2_\nu s^2 \right ) \xi^{}_{23} \right ] \; .
\end{equation}
It becomes obvious that ${\cal J}^{}_{\rm D} = 0$ will be a stable
result independent of the renormalization scales, provided
$\theta^{}_l$ or $\sin\phi$ initially vanishes at a given scale.
In comparison, we have
\begin{eqnarray}
\dot{\cal J}^{}_{\rm M} & = & \frac{Cy^2_\tau}{16\pi^2} \left \{
{\cal J}^{}_{\rm M} \left [ \left (c^2_\nu - s^2_\nu \right ) s^2
\left ( \zeta^{-1}_{12} c^2_{(\sigma -\rho)} + \zeta^{}_{12}
s^2_{(\sigma -\rho)} \right ) + \left ( c^2 - s^2_\nu s^2 \right )
\left ( \zeta^{-1}_{13} c^2_\rho + \zeta^{}_{13} s^2_\rho \right )
\right . \right . \nonumber \\
&  & \left . \left . ~~~~~ + \left ( c^2 - c^2_\nu s^2 \right )
\left ( \zeta^{-1}_{23} c^2_\sigma + \zeta^{}_{23} s^2_\sigma
\right ) \right ] + c^{}_\nu s^{}_\nu c s^2 \left ( C^{}_{12}
\widehat{\zeta}^{}_{12} + C^{}_{13} \widehat{\zeta}^{}_{13} +
C^{}_{23} \widehat{\zeta}^{}_{23} \right ) \right \} \;
\end{eqnarray}
for Majorana neutrinos, where
\begin{eqnarray}
C^{}_{12} & = & c^{}_l s^{}_l s^2 c^{}_\phi c^{}_{(\sigma -\rho)}
s^{}_{(\sigma -\rho)} \; ,
\nonumber \\
C^{}_{13} & = & \left [ c^{}_l s^{}_l c^{}_\phi \left ( s^2_\nu -
c^2_\nu c^2 \right ) + \left ( c^2_l - s^2_l \right ) c^{}_\nu
s^{}_\nu c \right ] c^{}_\rho s^{}_\rho \; ,
\nonumber \\
C^{}_{23} & = & \left [ c^{}_l s^{}_l c^{}_\phi \left ( c^2_\nu -
s^2_\nu c^2 \right ) - \left ( c^2_l - s^2_l \right ) c^{}_\nu
s^{}_\nu c \right ] c^{}_\sigma s^{}_\sigma \; .
\end{eqnarray}
One can see that ${\cal J}^{}_{\rm M}$ can be radiatively
generated from two non-trivial Majorana phases $\rho$ and
$\sigma$, even if it is initially vanishing at a specific scale.
Taking $\rho =\sigma =0$, we arrive at $C^{}_{12} = C^{}_{13} =
C^{}_{23} =0$ as well as $\dot{\rho} =\dot{\sigma} =0$. But it is
impossible to obtain the equality $\dot{\cal J}^{}_{\rm M}
(\rho=\sigma=0) = \dot{\cal J}^{}_{\rm D}$, because
$\zeta^{-1}_{12} = \xi^{}_{12}$, $\zeta^{-1}_{13} = \xi^{}_{13}$
and $\zeta^{-1}_{23} = \xi^{}_{23}$ (or equivalently $m^{}_1
m^{}_2 = m^{}_1 m^{}_3 = m^{}_2 m^{}_3 = 0$) cannot simultaneously
hold. This observation demonstrates again that the RG running
behavior of ${\cal J}^{}_{\rm M}$ is essentially different from
that of ${\cal J}^{}_{\rm D}$.

\section{Concluding remarks}

We have pointed out that the $\tau$-lepton dominance in the
one-loop RG equations of relevant neutrino mixing quantities
allows us to set a criterion for the choice of the most
appropriate parametrization of the lepton flavor mixing matrix
$U$: its elements $U^{}_{3i}$ (for $i=1,2,3$) should be as simple
as possible. Such a novel parametrization {\it does} exist, but it
is quite different from the ``standard" parametrization advocated
by the Particle Data Group and used in the literature. We have
shown that this parametrization can lead to greatly simplified RG
equations for three mixing angles and the physical CP-violating
phase(s), no matter whether neutrinos are Dirac or Majorana
particles.

The present work aims at the derivation of those one-loop RG
equations and some generic discussions about their important
features. A quantitative and detailed analysis of the RG running
behaviors of relevant neutrino mixing angles and CP-violating
phases is very desirable and will be done elsewhere. It is worth
emphasizing that our analytical results are so concise that they
can help understand radiative corrections to lepton flavor mixing
parameters in a much simpler and more transparent way. In
particular, they are expected to be very useful for model building
at a superhigh energy scale (e.g., the seesaw scale) to explore
possible flavor symmetries or flavor dynamics which are
responsible for the origin of neutrinos masses and leptonic CP
violation.

\vspace{0.5cm}

The author would like to thank S. Luo, J.W. Mei, H. Zhang and S.
Zhou for many helpful discussions. He is also grateful to H.
Fritzsch for warm hospitality at LMU of Munich, where this paper
was written. This work was supported in part by the National
Natural Science Foundation of China.

\end{document}